\documentclass[]{emulateapj}

\def\lya{Ly$\alpha$ }
\def\Msun{M_\odot}
\def\hMsun{h^{-1}M_\odot}
\def\hMpc{h^{-1}{\rm Mpc}}
\def\hkpc{h^{-1}{\rm kpc}}
\def\VhMpc{h^{-3}{\rm Mpc}^3}
\def\kms{{\rm km\, s^{-1}}}

\def\SBunit{{\rm erg\, s^{-1}\, cm^{-2}\, arcsec^{-2}}}

\slugcomment{Revised Draft, 20110704}
\shorttitle{Extended Lyman-Alpha Emission around Star-forming Galaxies}
\shortauthors{Zheng et al.}

\begin{document}

\title{
Extended Lyman-Alpha Emission around Star-forming Galaxies
}
\author{
Zheng Zheng\altaffilmark{1},
Renyue Cen\altaffilmark{2},
David Weinberg\altaffilmark{3},
Hy Trac\altaffilmark{4},
and
Jordi Miralda-Escud\'e\altaffilmark{5,6}
}
\altaffiltext{1}{Yale Center For Astronomy and Astrophysics, Yale University, 
                 New Haven, CT 06520; zheng.zheng@yale.edu
}
\altaffiltext{2}{Department of Astrophysical Sciences, Princeton University,
                 Peyton Hall, Ivy Lane, Princeton, NJ 08544
}
\altaffiltext{3}{Department of Astronomy, Ohio State University, Columbus, 
                 OH 43210
}
\altaffiltext{4}{Department of Physics, Carnegie Mellon University, Pittsburgh,
                 PA 15213
}
\altaffiltext{5}{Instituci\'o Catalana de Recerca i Estudis Avan\c cats,
                 Barcelona, Catalonia
}
\altaffiltext{6}{Institut de Ci\`encies del Cosmos, Universitat de Barcelona, 
                 Barcelona, Catalonia
}

\begin{abstract}
\lya photons that escape the interstellar medium of star-forming galaxies
may be resonantly scattered by neutral hydrogen atoms in the
circumgalactic and intergalactic media, thereby increasing the angular
extent of the galaxy's \lya emission. We present predictions of this extended,
low surface brightness \lya emission based on radiative transfer modeling in 
a cosmological reionization simulation. The extended emission can be detected 
from stacked narrowband images of \lya emitters (LAEs) or of Lyman break
galaxies (LBGs). Its average surface brightness profile has a central
cusp, then flattens to an approximate plateau beginning at an inner 
characteristic scale 
below $\sim$0.2 Mpc
(comoving), then steepens again 
beyond an outer characteristic scale of $\sim 1$ Mpc.  The inner scale marks 
the transition from scattered light of the central source to emission from 
clustered sources, while the outer scale marks the spatial extent of scattered 
emission from these clustered sources.  Both scales tend to increase with halo 
mass, UV luminosity, and observed \lya luminosity. The extended emission
predicted by our simulation is already within reach of deep narrowband
photometry using large ground-based telescopes.  Such observations would
test radiative transfer models of emission from LAEs and LBGs, and they would 
open a new window on the circumgalactic environment of high-redshift 
star-forming galaxies.
\end{abstract}

\section{Introduction}

High-redshift star-forming galaxies are becoming an important probe of galaxy 
formation, reionization, and cosmology. Ionizing photons of young stars in 
star-forming galaxies ionize neutral hydrogen atoms in the interstellar medium 
(ISM), and each subsequent recombination has a probability of $\sim 2/3$ of
ending up as a \lya photon \citep{Partridge67}. After escaping the ISM, these 
\lya photons can be scattered by neutral hydrogen atoms in 
the circumgalactic and intergalactic media (IGM), which tends to make the \lya 
emission extended. In this paper, we present predictions for the extended \lya 
emission associated with high-redshift star-forming galaxies from a radiative 
transfer model of \lya emission applied to a hydrodynamic cosmological 
simulation \citep{Zheng10}.

As a result of reprocessed ionizing photons, prominent \lya emission can be a 
characteristic of star-forming galaxies, which can be used to detect 
high-redshift galaxies. Galaxies detected through the strong \lya emission 
associated with them (e.g., from narrowband photometry) are dubbed Lyman-alpha 
emitters (LAEs) and an increasing number of such galaxies have been
discovered \citep[e.g.,][]{Hu96,Cowie98,Rhoads03,Malhotra04,Gawiser07,
Ouchi08,Guaita10,Ouchi10}. High-redshift star-forming galaxies can also be 
detected from broadband photometry with Lyman break technique, which are 
termed Lyman break galaxies (LBGs; e.g., \citealt{Steidel03}). As long as 
\lya photons reprocessed from ionizing photons can escape from the ISM, \lya 
emission is also expected to be associated with LBGs. \lya emission encodes 
useful information about star-forming galaxies and their environments, 
such as star formation rate (e.g., \citealt{Madau98}), kinematics of 
ISM gas (e.g., \citealt{Steidel10}), and neutral fraction of IGM gas 
(e.g., \citealt{Dijkstra07}). \lya photons usually experience complex radiative
transfer processing (resonant scatterings) in media with neutral hydrogen 
atoms, which complicates the interpretation of observed \lya emission 
properties.

\citet{Zheng10} (hereafter Paper I) present a physical model of \lya emission 
from LAEs by solving \lya radiative transfer in the circumgalactic and 
intergalactic media.  The \lya radiative transfer calculation is performed in 
a cosmological volume ($100\hMpc$ on a side) from a state-of-the-art 
radiation-hydrodynamic reionization simulation \citep{Trac08}. The calculation 
uses a $768^3$ grid to sample the density, velocity, and temperature of the 
neutral hydrogen gas in the simulation and is applied to all the sources 
residing in halos above $5\times 10^9\hMsun$. Resonant 
scatterings enable \lya photons to probe the circumgalactic and intergalactic 
environments (density and velocity structures) around star-forming galaxies.
This leads to a coupling between the observed \lya emission properties and the 
environments. This simple physical model is able to explain an array of 
observed properties of $z\sim$5.7 LAEs the Subaru/{\it XMM-Newton}
Deep Survey (SXDS; \citealt{Ouchi08}), including \lya spectra, morphology, 
apparent \lya luminosity function (LF), shape of the ultraviolet (UV) LF, and 
the distribution of \lya equivalent width. The selection imposed by
the environment dependent radiative transfer also introduces interesting new
features in the clustering of LAEs (\citealt{Zheng11}, hereafter Paper II).

In the above radiative transfer model, while the number of \lya photons is 
conserved after they escape the ISM, the scatterings in the circumgalactic and 
intergalactic media cause the \lya emission to spread spatially. Therefore, 
one generic prediction of the model is an extended \lya emission halo around 
a star-forming galaxy. Observationally, only a fraction of \lya photons can be 
detected for an individual source, those included in the central part of the 
extended \lya emission with high enough surface brightness (tip of the 
iceberg). The outskirts of the \lya halo with low surface brightness is 
typically buried in the sky noise. 

In this paper, we show that it is possible to detect the bottom of the 
iceberg by stacking the narrowband images of a large number of sources to 
suppress the sky noise. We present the predictions of the extended \lya emission
in the stacked image from our radiative transfer model and discuss what we can 
learn from it. Our radiative transfer modeling is performed for sources at 
$z\sim 5.7$. We study properties of \lya surface brightness profile from 
stacked images in Section 2. In Section 3, we discuss the observational 
prospects. We summarize our results in Section 4.

Throughout the paper, we adopt a spatially flat $\Lambda$CDM cosmological model
for our calculations, with a matter density parameter $\Omega_m=0.28$ and a
Hubble constant $h=0.70$ in units of $100\, \kms {\rm Mpc}^{-1}$. Distances are
expressed in comoving units unless mentioned clearly otherwise.

\begin{figure}
\plotone{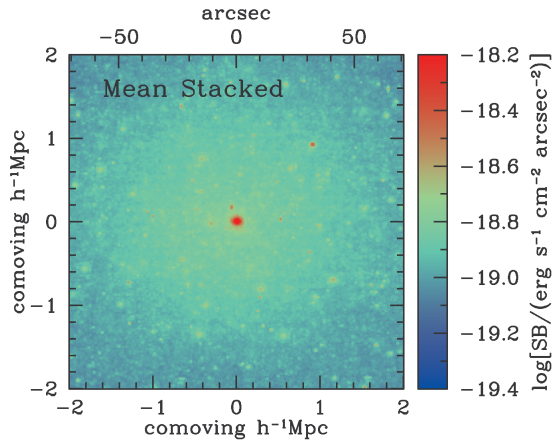}
\caption[]{
\label{fig:stack_img}
Stacked narrowband \lya image for sources residing in halos of 
$10^{11}\hMsun$ at $z\sim$5.7. A mean algorithm is used to stack
$\sim$440 sources.
}
\end{figure}

\section{Extended \lya Emission around Star-forming Galaxies}
\label{sec:extended}

The \lya radiative transfer calculation in Paper I is performed in a simulation
box of 100$\hMpc$ (comoving) on a side with the neutral hydrogen density, 
temperature, and peculiar velocity sampled with a 768$^3$ grid. The cell size 
is therefore 130.2$\hkpc$ (comoving), about the virial diameter of a 
$1.8\times 10^{10}\hMsun$ halo. The star-forming region in a high-redshift 
galaxy, where \lya photons in our study originate, is only a few kpc (comoving)
across \citep[e.g.,][]{Taniguchi09}. So the initial \lya emission is treated 
as a point source at the halo center for the purpose of the radiative transfer 
calculation.
The size of the pixel for collecting \lya 
photons is 8 times finer, corresponding to 16.3$\hkpc$ (comoving) or 
0.58\arcsec. With the cell resolution, the gas distribution is uniform 
inside the virial radius of small halos (below a few times $10^{10}\hMsun$),
which would have some effect on the \lya surface brightness profile. 
However, since \lya photons are initially emitted from a point source and 
the dynamics of the infall region around halos plays a significant role in
determining the distribution of \lya photons (Paper I), the resolution we use
is sufficient for obtaining the generic features in the extended \lya 
emission. We present a resolution test in \S~2.5, and the main results 
presented in this paper would remain valid with improved resolution.
Our calculation does not address the effects of galactic wind and dust, and 
we discuss these model uncertainties in \S~2.4 and \S~3.

\subsection{Stacked \lya Image and \lya Emission at Fixed Halo Mass}

We start from the stacked \lya image for sources residing in halos of fixed
mass, $10^{11}\hMsun$. In our model, the intrinsic \lya luminosity, which is
the the total amount of \lya emission from the reprocessed ionizing photons, 
is tightly correlated with the halo mass, therefore this stacking is basically 
for sources at fixed intrinsic \lya luminosity or UV luminosity. 

Our radiative transfer model produces a three-dimensional array, recording the
\lya spectra as a function of spatial position on the sky. We construct the 
narrowband \lya image from this array, with a filter width similar to  
that in the $z\sim 5.7$ SXDS (see Paper I). Our simulation box has about the 
same area as SXDS with a redshift depth three times larger. So we are able
to construct narrowband \lya images for three SXDS-like fields.
Each image corresponds to an ideal
case with perfect continuum and sky subtraction. Examples of images of 
individual sources can be found in Paper I (Fig.4 and Fig.5).
We stack all the source images together as would be done with the narrowband
observation. For each source, the image includes the center and surrounding 
pixels (up to a radius of $\sim$ 10$\hMpc$). The center pixel is chosen to 
be the one that contains the halo center. In other words, the center 
corresponds to that in the UV band. It is evident that the surrounding pixels 
can include \lya photons from other neighboring sources. For all the sources in
a narrow bin (0.16 dex) of halo mass around $10^{11}\hMsun$ (about 440 sources
for each of the SXDS-like fields), we stack their images with their centers 
aligned. 

\begin{figure}
\plotone{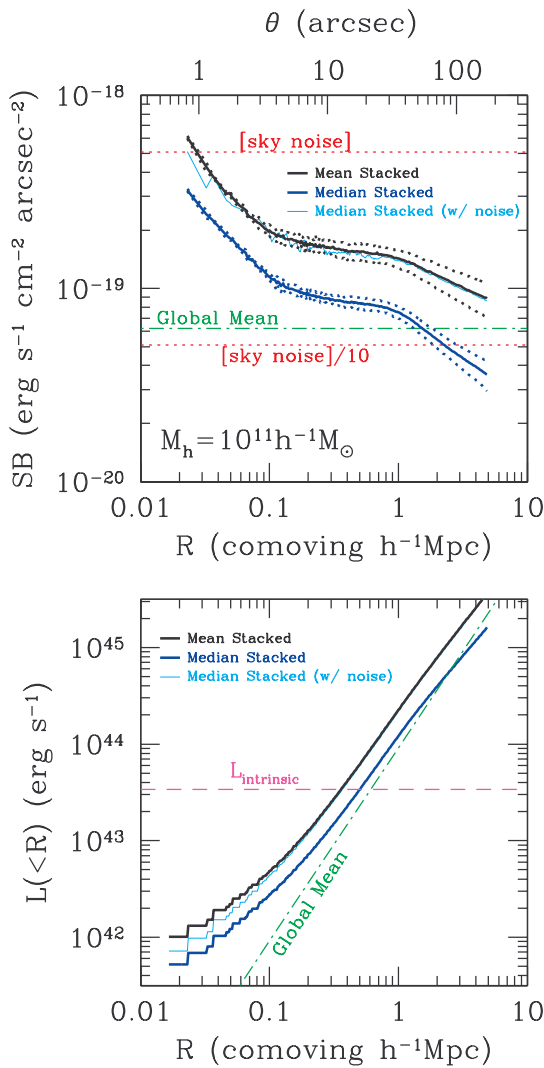}
\caption[]{
\label{fig:SB_M11}
Surface brightness profile of extended \lya emission from stacked narrowband 
image for sources residing in halos of $10^{11}\hMsun$. Top panel: the two 
thick solid curves are from mean and median stacked noise-free images, 
respectively. The associated dotted curves indicate the scatter around the 
mean surface brightness, based on stacked images from three SXDS-like fields 
in our model. 
The thin solid curve is from the {\it median} stacked noise-added images, where
the Poisson photon noise (source + sky) is computed for a 4-hour exposure 
with the Subaru telescope.
The upper horizontal dotted line is the surface brightness level corresponding 
to the 1$\sigma$ sky noise in a $2\arcsec$ diameter aperture, assuming a 4-hour
exposure with the Subaru telescope. It drops to the lower horizontal dotted 
line if 100 sources are stacked. The horizontal dot-dashed line is the global 
mean \lya surface brightness if the \lya flux from all the sources in each 
field is uniformly distributed across the field. Bottom panel:
The cumulative \lya luminosity distributions from the stacked images. The 
dashed line denotes the intrinsic \lya luminosity of one source residing in
a halo of $10^{11}\hMsun$. The dot-dashed line is the profile from the global
mean surface brightness profile, which is $\propto R^2$.
}
\end{figure}

Typically there are two algorithms to generate a stacked image, taking 
either the mean or the median.
The stacked \lya image from the mean algorithm is shown in 
Figure~\ref{fig:stack_img}. The mean algorithm has the advantage of 
conserving the flux, even though it may be affected by outliers in the flux
distribution, showing up as many small clumps in the stacked image. The 
median stacked image has the advantage of insensitivity 
to outliers. We find that at a fixed projected radius, the observed flux 
roughly follows a log-normal distribution among different sources, which
causes the surface brightness from the median stacked image to be lower 
in amplitude than that from the mean stacked image (see solid curves in 
Figure~\ref{fig:SB_M11}). The stacking cases mentioned above 
correspond to an ideal situation of no photon noise. When total photon noise
(source + sky) is added, the surface brightness profile from the mean 
stacked image remains the same and that from the median stacked image 
would change. In the limit of large noise (compared to the signal 
distribution), the flux distribution is completely modified by the noise
distribution, and the median value is driven to be close to the mean (also see
\citealt{Pieri10}). 
In practice, the large noise limit is typical at large radii for the case 
considered here (see below), so the surface brightness profile from the 
median stacked image approaches that from the mean stacked one. See the
thin solid curve in Figure~\ref{fig:SB_M11}.
In what 
follows, we focus on presenting our results from the mean stacked image. We 
compute the surface brightness profile by averaging the flux in annuli at 
different radii, therefore the effect of the grainy feature seen in the image 
is greatly reduced.

The surface brightness profile from the stacked image shows a few interesting 
features. It has two characteristic scales that are associated with steep 
changes in the profile slope. The inner slope change occurs at 
$R_{\rm in}\sim 0.1\hMpc$ (possibly upper limit, see \S~\ref{sec:resolution})
and the outer one at $R_{\rm out}\sim 1\hMpc$. 
Inside the inner characteristic radius $R_{\rm in}$, the profile appears as a 
central cusp, roughly following a steep power-law. Between the inner and outer 
characteristic radii, $R_{\rm in}<R<R_{\rm out}$, the surface brightness 
decreases slowly with increasing radius, close to a plateau. Beyond the outer 
characteristic radius $R_{\rm out}$, the surface brightness profile steepens 
and drops, forming an extended tail. As shown later, the two scales have clear 
physical meanings.

With the mean surface brightness profile, we compute the cumulative luminosity 
as a function of projected radius (bottom panel of Figure~\ref{fig:SB_M11}).
It turns out that the luminosity from the stacked image does not converge at 
large radius. Because of the plateau in the surface brightness profile, the 
luminosity profile approximately follows $R^2$, a trend similar to that from 
the global mean surface brightness (dot-dashed line). The global mean profile
is computed by uniformly distributing the total \lya flux in the original 
narrowband image across the whole image. The plateau in the stacked
surface brightness profile has a much higher amplitude than the global mean. 
As we show later, the higher signal is caused by the clustered 
neighboring sources.

In the top panel of Figure~\ref{fig:SB_M11}, we also show the level of sky 
noise (dotted lines). The calculation is based on a 4-hour observation with
the NB816 filter using the Subaru telescope\footnote{We use the Subaru 
Imaging Exposure Time Calculator at http://www.naoj.org/cgi-bin/img\_etc.cgi.}, which is roughly the setup for the survey of $z\sim 5.7$ LAEs in SXDS 
\citep{Ouchi08}. In the NB816 band, the sky (AB magnitude of 
20.4mag/arcsec$^2$ in a dark night) has a surface brightness of 
$f_{\rm sky} \sim 1.36\times 10^{-15}\SBunit$. 
The mean sky photon count in a 2\arcsec diameter aperture from a 4-hour 
exposure with the Subaru telescope reaches $N_{\rm sky}\sim 7.1\times 10^6$. 
Therefore, the 1$\sigma$ sky noise in a 2\arcsec diameter aperture corresponds 
to a surface brightness level of $f_{\rm sky}/\sqrt{N_{\rm sky}} 
\sim 5.1\times 10^{-19}\SBunit$,
which is higher than the signal of 
interest on large scales (Fig.~\ref{fig:SB_M11}). The noise per pixel is 
expected to be suppressed by $\sqrt{N}$ with $N$ sources stacked. The 
suppression works even better for detecting the signal at large radii, because 
the annuli used to compute the mean surface brightness profile can be much 
larger than the 2\arcsec aperture. Within each annulus, however, the noise
in different pixels is correlated, since the same pixel from the 
original image may fall into the annulus several times from centering 
different sources. The correlations need to be accounted for in estimating
the noise level at each radius for the stacked profile. In practice, given the 
low surface brightness of the extended \lya emission and the high sky noise 
level, stacking the image before subtracting the continuum and sky may help 
to reduce errors caused by such subtraction and to increase the sensitivity.

\subsection{Decomposition of the \lya Surface Brightness Profile in the Stacked Image}

\begin{figure}
\plotone{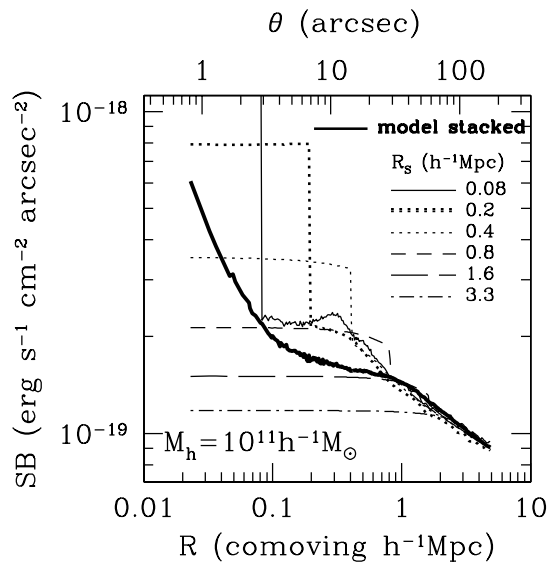}
\caption[]{
\label{fig:stack_tophat_M11}
\lya surface brightness profile from (mean) stacked images with different 
assumptions about the extent of \lya emission. Sources considered here reside 
in halos of $10^{11}\hMsun$. The thick solid curve is the prediction from the 
radiative transfer model, which is the same as that in Fig.~\ref{fig:SB_M11}.
Other curves are for cases with no radiative transfer applied but with the \lya 
emission from a source artificially set to spread into a uniform ``seeing'' 
disk of radius $R_s$, which are to illustrate the effect of source extent 
combined with source clustering. 
The signal outside of this radius comes from the neighboring sources as a 
consequence of clustering. See the text.
}
\end{figure}

To understand the features in the stacked \lya surface brightness profile and 
the origin of the two characteristic scales, we 
perform a test by assigning artificial surface brightness profiles for 
individual sources. The image of each source (in halos above 
$5\times 10^9\hMsun$) is assumed to be a circular disk of radius $R_s$ with
uniform surface brightness normalized such that the total luminosity is 
the intrinsic \lya luminosity from
the source. That is, we replace the intrinsic point source with a uniform
``seeing'' disk or (equivalently) modify it by a tophat point spread function 
(PSF). We then follow the same procedure as above to form the stacked
image for sources in $10^{11}\hMsun$ halos and derive the \lya surface 
brightness profile. No radiative transfer is performed in this artificial 
model, and we use it to illustrate the effect of source extent
combined with source clustering.

Figure~\ref{fig:stack_tophat_M11} shows the series of surface brightness 
profiles with different ``seeing'' disk radii $R_s$. For a small ``seeing''
disk radius
($R_s=0.08\hMpc$), sources are close to point-like. The central spike (inside
$R_s$) is clearly seen in the profile from the stacked image. Outside
of this radius, we still see signals, which are obviously contributed from 
neighboring sources as a result of clustering. 

The surface brightness profile has contributions from 
the central sources and from the neighboring clustered sources, analogous to
the case of weak gravitational lensing around galaxies.
In the spirit of the halo model \citep[e.g.,][]{Seljak00,Berlind02,Cooray02},
the profile can be decomposed into two components, that from the individual
source (the one-halo term) and that caused by the clustering (the two-halo
term). Formally, we can write the ``two-halo'' term of the stacked surface 
brightness profile $\Sigma_{\rm 2h}({\mathbf R})$ as
\begin{eqnarray}
\hspace{-0.8cm}
\label{eqn:2halo}
\Sigma_{\rm 2h}({\mathbf R}|M) & = &
           \int dM^\prime \frac{dN}{dM^\prime} \int d^2{\mathbf R^\prime} \nonumber \\
       & &    \left[1+w({\mathbf R^\prime}|M,M^\prime)\right]
           \Sigma_{\rm 1h}({\mathbf R}-{\mathbf R^\prime}|M^\prime),
\end{eqnarray}
where $dN/dM^\prime$ is the surface number density of halos, $w$ is the 2D 
two-point
correlation function (essentially the angular correlation function), and
$\Sigma_{\rm 1h}$ is the surface brightness profile of individual sources
(the one-halo term). For simplicity, equation~(\ref{eqn:2halo}) assumes that 
$\Sigma_{\rm 1h}$ depends only on halo mass, while in reality it 
depends on the environment around halos as well (see below). We emphasize that
the one-halo term in the stacked profile is defined as the \lya emission from 
the central source alone (i.e., contributed from all the sources to be 
center-aligned and stacked). Even though neighboring sources can have their 
\lya photons scattered into the one-halo scales of the central source, these 
photons still belong to the contribution from the two-halo term.

In the limit of compact sources (small $R_s$ in the uniform ``seeing'' disk 
case), 
$\Sigma_{\rm 1h}$ can be replaced with 
$L(M^\prime)\delta_D({\mathbf R}-{\mathbf R^\prime})$, where $L(M^\prime)$
is the intrinsic \lya luminosity and $\delta_D$ is the Dirac delta function. 
The two-halo term then becomes 
\begin{equation}
\label{eqn:2halo_pointsource}
\Sigma_{\rm 2h}(R|M)=
           \int dM^\prime \frac{dN}{dM^\prime} L(M^\prime)
           \left[1+w(R|M,M^\prime)\right].
\end{equation}
For the small ``seeing'' disk case in Figure~\ref{fig:stack_tophat_M11}
($R_s=0.08\hMpc$), the flattening below $0.3\hMpc$ is a sign of the halo 
exclusion effect, which leads to a zero signal in the 3D 
two-point correlation function and a constant in the angular two-point 
correlation function $w$.\footnote{
The bump around $0.3\hMpc$ can be understood as follows. Roughly speaking, 
$w(R)\propto \int_{Z_0}^\infty \xi(R,Z) dZ$ with 
$Z_0=\sqrt{r_c^2-R^2}$ in the regime that $R<r_c$, where 
$r_c\sim 0.2$--0.3$\hMpc$ is the halo exclusion scale and $\xi$ is the 3D 
two-point correlation function (zero on scales below $r_c$). As $R$ increases, 
the competition between a decreasing $\xi$ on average and an increasing path 
length of integration (i.e., a decreasing lower limit $Z_0$) leads to a small 
bump in $w$ around $0.3\hMpc$.
If the seeing disk is comparable or larger than the size of the halo 
(e,g., the $R_s=0.2\hMpc$ case in Figure~\ref{fig:stack_tophat_M11} for 
$10^{11}\hMsun$ halos), the bump is smeared out.
}

According to Equation~(\ref{eqn:2halo_pointsource}),
we expect to see a flattened profile on large scales, where $w\ll 1$. On 
these scales, which are out of the range of the plot, it is hard in practice 
to disentangle such a uniform background from the sky background. We therefore
limit ourselves to the regime where $w$ is still large.

By varying the radius $R_s$ of the uniform ``seeing'' disk, we show the effect 
of the spatial extent of the \lya emission on the surface brightness profile 
from the stacked image (Figure~\ref{fig:stack_tophat_M11}). As the ``seeing''
disk size increases, the two-halo term of the profile is smoothed on scales 
below the ``seeing'' disk size, which results in a decrease in the amplitude. 
In our setup, the uniform ``seeing'' disk has a surface brightness of 
$1.4\times 10^{-19}(R_s/0.4\hMpc)^{-2}\SBunit$ for $R<R_s$, which is by 
definition the one-halo term contribution. However, the small-scale
surface brightness (in the one-halo regime) in 
Figure~\ref{fig:stack_tophat_M11} decreases much slower
than $R_s^{-2}$, as a consequence of the contribution from the smoothed 
two-halo term. For $R_s=1.6\hMpc$, the signal is already dominated by the 
two-halo term on all scales. The approximate plateau 
seen in the stacked profile from the radiative transfer 
model (thick solid curve) is close to the case with a ``seeing'' disk size of
$R_s=1.6\hMpc$. The test suggests that the plateau feature between the two
characteristic scales mainly originates 
from spatially extended, scattered \lya emission from clustered sources around 
the central source.   

\begin{figure}
\plotone{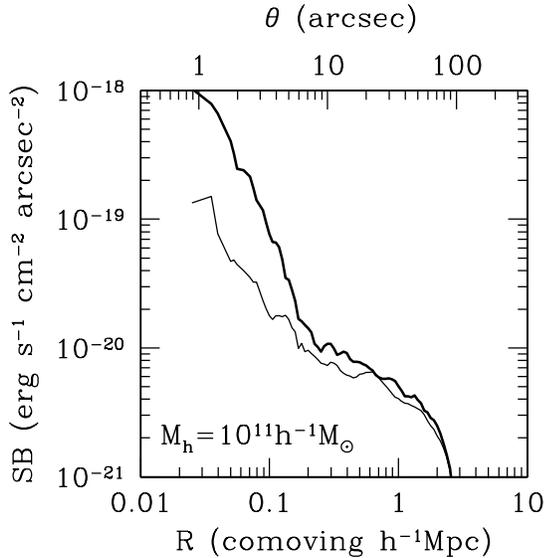}
\caption[]{
\label{fig:PSF}
Mean \lya surface brightness profile from an individual source. The host halo
mass for this source is $10^{11}\hMsun$. The thick and thin curves are the
same source viewed along different directions. 
}
\end{figure}

The uniform ``seeing'' disk only gives us a rough idea on the features in the 
extended \lya emission. To be more accurate, we need to examine the \lya 
surface brightness profile of individual sources. 

As an example, we show in Figure~\ref{fig:PSF} the surface brightness profiles 
of one individual source in the simulation box from the radiative transfer 
model. The two curves correspond to the profiles viewed along different 
directions, and the difference reflects the environment dependent \lya 
radiative transfer, as discussed in detail in Paper I. The surface brightness 
has a steep drop below $0.3\hMpc$. Then it levels off until a cutoff is 
reached around 2--3$\hMpc$. The inner surface brightness profile of one 
source roughly goes as $R^{-2}$ and the luminosity increases logarithmically
with radius. Most of the luminosity comes from near the cutoff radius at 
$\sim$2$\hMpc$, which is the reason why the plateau up to such a scale in 
the total profile is produced (see below).

As discussed in Paper I, the shape of the surface brightness profile largely 
depends on the velocity field. Below $0.3\hMpc$, the contribution to the 
surface brightness profile mainly
comes from the \lya photons that are last scattered in the infall region 
around the halo. On larger scales, where the profile levels off, the last
scattering of \lya photons occurs in the IGM that undergoes Hubble flow. The 
cutoff radius corresponds to those photons that are the bluest after 
escaping the infall region. They can travel far away from the source before
being redshifted into the line center to be significantly scattered in the IGM.
The exact shape of the surface brightness profile also depends on the initial
frequency of \lya photons (see \S~\ref{sec:line_prof}). 

\begin{figure}
\plotone{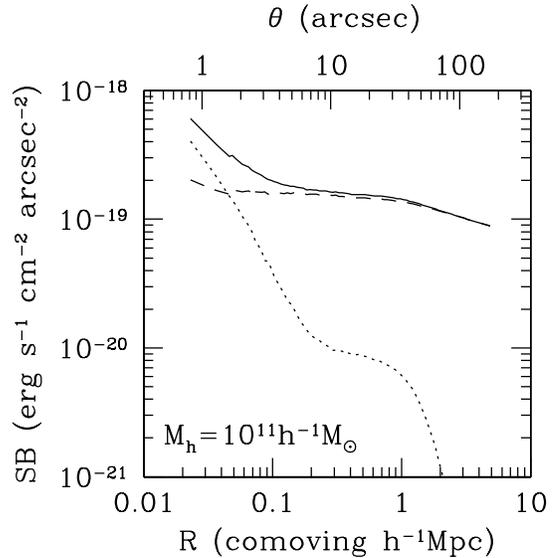}
\caption[]{
\label{fig:stackSB_1h2h}
One-halo and two-halo contributions to the \lya surface brightness profile.
The dotted curve is the average profile from individual sources (one-halo 
term), and the dashed curve is the profile from source clustering (two-halo 
term). The solid curve is the sum of the two contributions, which would be the 
one from the stacked image.
}
\end{figure}

As emphasized in Paper I and Paper II, \lya radiative transfer depends on
circumgalactic and intergalactic environments (density and velocity 
structures). Even for sources in halos of the same mass, there is a broad 
distribution of environments. Therefore, the one-halo term of the surface
brightness profile is not identical for sources in halos of fixed 
mass or for the same source viewed along different directions 
(Figure~\ref{fig:PSF}). The one-halo term in Equation~(\ref{eqn:2halo}) should
have dependence on environment in addition to halo mass. To obtain the mean 
one-halo term for the stacked image and decompose the surface brightness 
profile, we perform \lya radiative transfer calculations separately for each 
source with halo mass in the $10^{11}\hMsun$ mass bin. We then stack the 
individual one-halo terms to obtain the mean. 

The dotted curve in Figure~\ref{fig:stackSB_1h2h} is the mean one-halo term
for sources in $10^{11}\hMsun$ halos. The shape is similar to that seen in 
Figure~\ref{fig:PSF}. The dashed curve is the two-halo term, inferred from 
subtracting the one-halo term from the total surface brightness profile.
The two-halo term can be expressed as a convolution (Equation~\ref{eqn:2halo}),
with the convolution kernel being the one-halo term of the clustered 
star-forming halos of {\it all} masses. The shape of the one-halo term at 
different halo mass is similar to that shown in Figure~\ref{fig:stackSB_1h2h}. 
In the two-halo term, a plateau reaches out to a radius of 1--2$\hMpc$, which
is around the cutoff scale of the mass-averaged one-halo term. This
is not a coincidence. The scale where the plateau in the two-halo term ends 
marks the spatial extent of the extended \lya emission of the clustered 
sources, since on scales larger than this the smoothing effect from the source 
PSF is small. As the two-halo
term dominates on large scales, the above scale is just the outer 
characteristic scale $R_{\rm out}$ seen in the total surface brightness 
profile. The inner characteristic scale $R_{\rm in}$ marks the transition from 
the one-halo term dominated regime to the two-halo term dominated regime.
The realistic decomposition of the surface brightness profile confirms the 
results of the test in Figure~\ref{fig:stack_tophat_M11} that the origin of 
the plateau between the two characteristic scales in the total surface 
brightness profile is the extended \lya emission.

\begin{figure}
\plotone{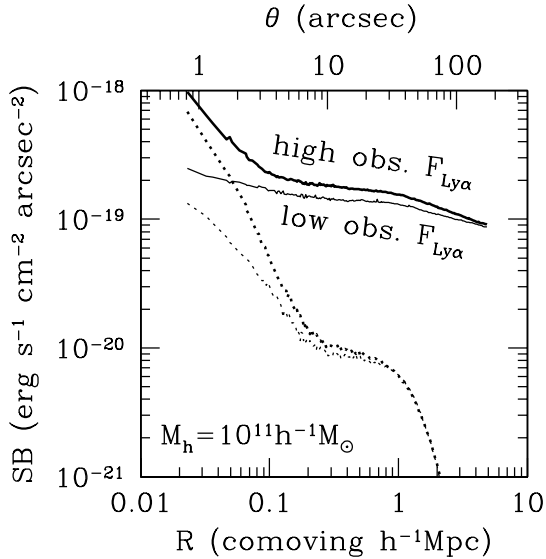}
\caption[]{
\label{fig:stackSB_halfhalf}
Surface brightness profile of extended \lya emission from stacked narrowband
image as a function of observed \lya luminosity. All sources reside in halos 
of $10^{11}\hMsun$. Thin (thick) curves are for the faint (bright) half of the
sources based on the observed \lya luminosity, which is contributed by the 
central, high surface brightness region. Dotted curves are the corresponding
one-halo profiles. See the text for details.
}
\end{figure}

\begin{figure*} 
\plottwo{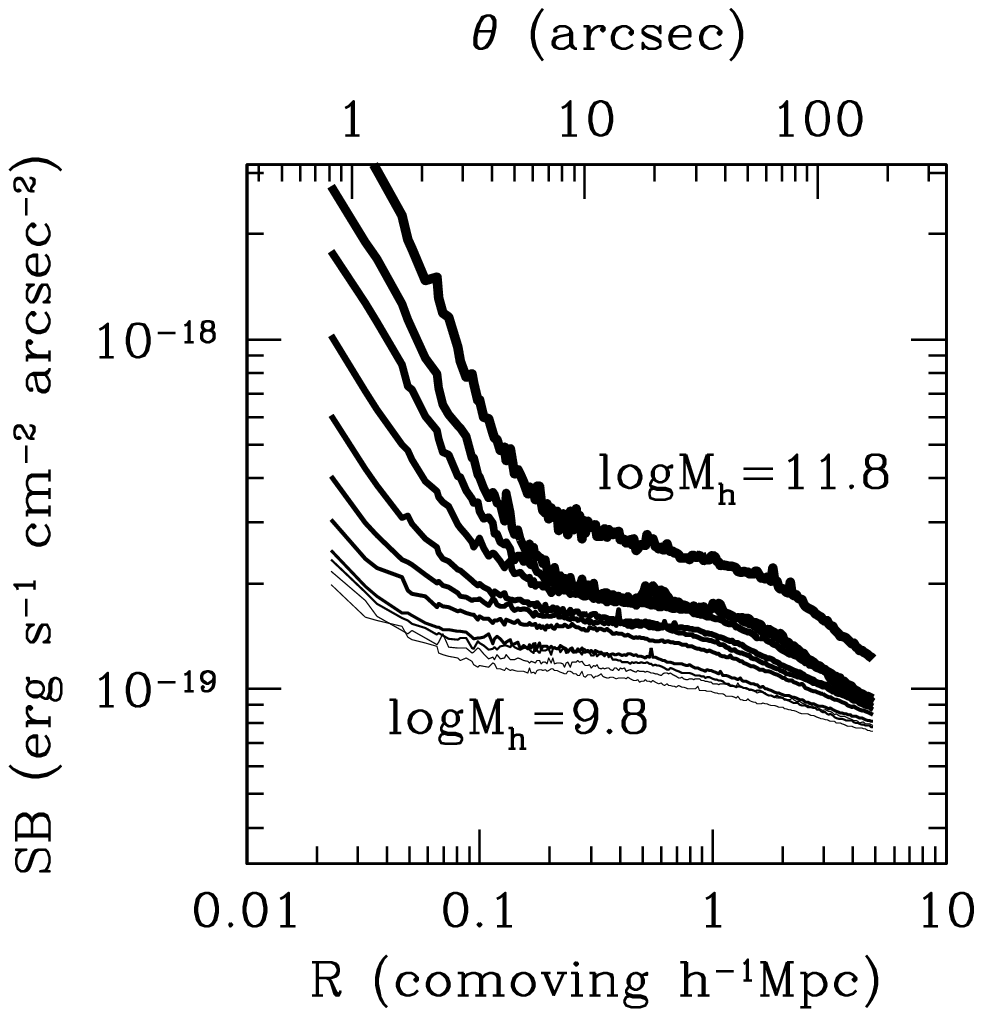}{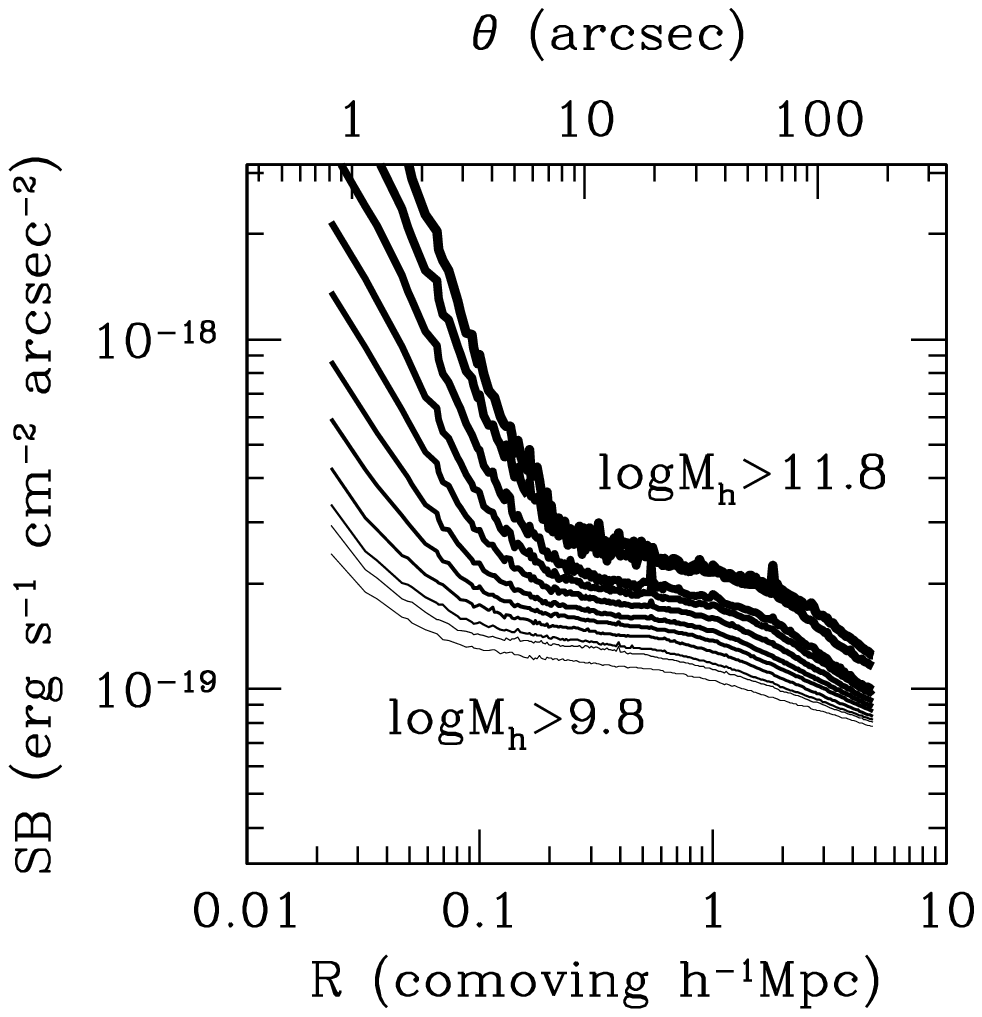}
\caption[]{
\label{fig:SB_Mh}
Dependence of mean \lya surface brightness profile on host halo mass (or UV
luminosity). {\it Left:} Profiles for samples of galaxies in different bins
of halo mass. From bottom to top curves, host halo mass increases from 
$\log[M/\hMsun]=9.8$ to 11.8, with an interval of 0.2 dex between two adjacent
curves. {\it Right:} Similar to the left panel, but for samples of galaxies
in halos above different mass thresholds.
}
\end{figure*}

The environment dependent \lya radiative transfer couples the observed \lya 
emission with the circumgalactic and intergalactic environments. Many
consequences of such a coupling are studied in Paper I and Paper II. For the
stacked surface brightness profile, the environment dependence also appears, 
as detailed below. 

At fixed halo mass, which corresponds to fixed intrinsic \lya luminosity or
UV luminosity in our model, the observed \lya luminosity has a broad 
distribution, reflecting the distribution of environments. In our model, for
each source we link together the pixels above a surface brightness threshold 
similar to that used in \citet{Ouchi08} to define the observed emission, 
and if none of the pixels around the central source exceeds the threshold, 
we simply take the flux in the pixel at the source position as the observed 
emission (see Paper I for details). The observed luminosity is only a fraction 
of the extended \lya emission, coming from the central, high surface 
brightness part of the source. The one-halo term of the surface brightness 
profile is expected to be closely related to the environment. 

We divide the sources into two samples of equal size, according to the 
observed \lya luminosity. In Figure~\ref{fig:stackSB_halfhalf}, we show the 
stacked surface brightness profile for the bright half and faint half, 
respectively. The dotted curves show the corresponding one-halo terms. 
We note that the fluxes inferred from the one-halo terms of the two cases
are not necessarily the same. The reason is that the scattered emission is 
not isotropic, being sensitive to the density and velocity structures around 
the source. Overall, the sample of sources with lower observed \lya luminosity 
has a shallower stacked surface brightness profile at small radius and there
is no clear scale of the transition from the inner cusp to the approximate 
plateau. The result implies that the stacked \lya narrowband image for 
star-forming galaxies with weaker observed \lya emission would appear to be 
less compact. 

\subsection{Dependence on UV and \lya Luminosity}

So far we have focused on the \lya surface brightness profile for sources in 
halos
of $10^{11}\hMsun$. The profile is expected to be a function of the halo mass.
We show the dependence on halo mass in Figure~\ref{fig:SB_Mh}, both for halo
in narrow mass bins (left panel) and above different mass thresholds (right 
panel). Since in our
model the UV luminosity is tightly correlated with halo mass, the plot also
represents a sequence of \lya profiles as a function of UV luminosity. 

The amplitude of the stacked profile increases with halo mass. On small scales 
this increase is largely a reflection of the fact that halos of higher
masses host sources of higher \lya luminosity, while on large scales there
is an additional contribution from stronger source clustering around more
massive halos.

On small scales (inside the inner characteristic radius), where the one-halo 
profile makes a substantial contribution, sources in higher mass halos show a 
steeper profile. The inner characteristic scale $R_{\rm in}$, where the 
profile starts to level off, increases with halo mass. It implies that sources 
in higher mass halos have more extended \lya emission. The outer characteristic
scale $R_{\rm out}$, where the profile starts to steepen again,  
also increases with halo mass.
As discussed above, this scale is an indication of the spatial extent of the 
\lya emission of the clustered sources. Massive halos are typically 
located in dense environments, therefore the clustered sources around them 
tend to be in massive halos as well. On scales larger than $R_{\rm out}$, 
the extended tail of the profile for higher 
mass halos is steeper. This is also a consequence of the higher clustering 
amplitude, since the surface brightness profile is proportional to (1+$w$) as 
in Equation~(\ref{eqn:2halo}), and the unity factor tends to flatten the 
profile for weakly clustered sources. 

Overall, the stacked \lya surface brightness profile is steeper for sources
in higher mass halos, and the two characteristic scales increase with halo 
mass. On small scales where the central cusp dominates, if we approximate the 
profile by a power law, we find that the power-law index decreases with halo 
mass, ranging from $-0.5$ to $-1.4$ (from 
$-0.5$ to $-1.7$) for the mass bin (threshold) samples shown in 
Figure~\ref{fig:SB_Mh}.

The halo mass (UV luminosity) sequence does not have a priori selection based 
on observed \lya emission, so it applies to the case of LBGs.

\begin{figure*}
\plottwo{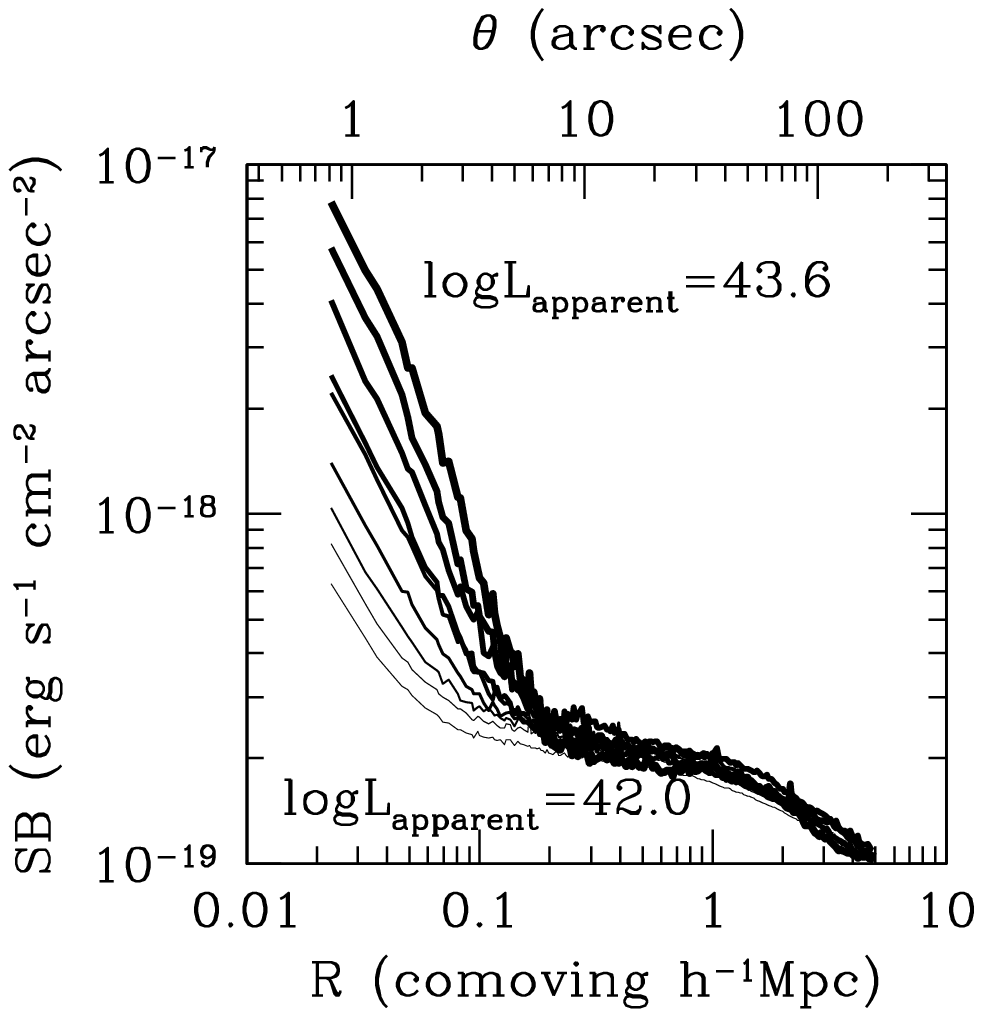}{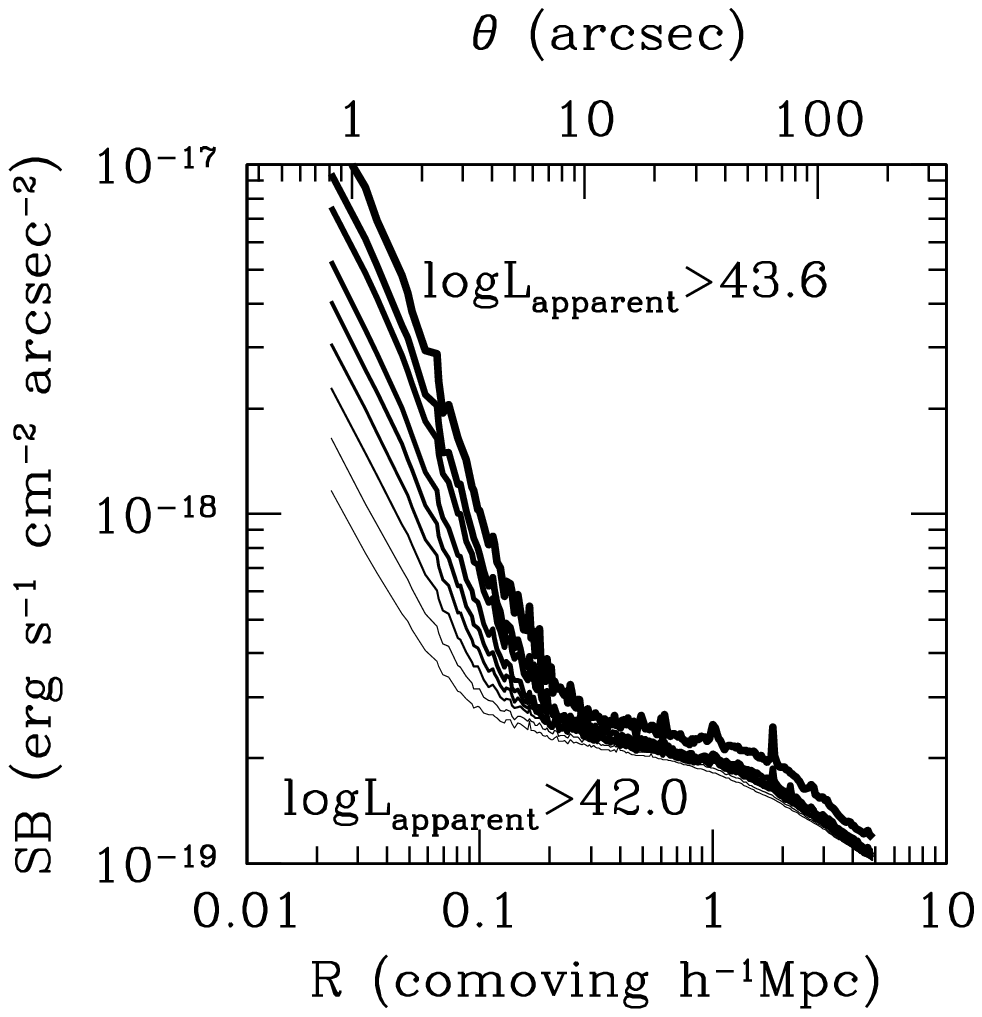}
\caption[]{
\label{fig:SB_Lap}
Dependence of mean \lya surface brightness profile on the apparent (observed) 
\lya luminosity. {\it Left:} Profiles for samples of galaxies in different bins
of apparent \lya luminosity. From bottom to top curves, apparent \lya 
luminosity increases from $\log[L_{\rm apparent}/({\rm erg\, s^{-1}})]=42.0$ 
to 43.6, with an interval of 0.2 dex between two adjacent
curves. {\it Right:} Similar to the left panel, but for samples of galaxies
above different thresholds in apparent \lya luminosity.
}
\end{figure*}

In practice, the brightness profile that can be easily measured in a 
narrowband survey is the stacked image of the identified LAEs, classified 
according to their apparent (observed) \lya flux. 
Figure~\ref{fig:SB_Lap} shows the stacked \lya surface
brightness profile for LAEs as a sequence of the apparent (observed) \lya
luminosity, for both luminosity-bin samples (left panel) and 
luminosity-threshold samples (right panel).  

The apparent \lya luminosity couples to the circumgalactic and intergalactic 
environment around sources as a result of the \lya radiative transfer, which
imposes a strong selection effect for sources that are identified as LAEs.
On small scales (inside $R_{\rm in}$), the stacked profile for LAEs appears 
to be steeper than those seen in cases in Figure~\ref{fig:SB_Mh} for sources 
without \lya selection. In Figure~\ref{fig:SB_Lap}, the central cuspy profile 
has a slope of $-0.9$ -- $-1.8$ ($-1.1$ -- $-1.8$) for luminosity bin 
(threshold) samples, with steeper slopes for higher observed \lya luminosities.
The amplitude increases with the apparent \lya luminosity. On scales above
the inner characteristic scale, the amplitude also increases with \lya 
luminosity, but the dependence is much weaker. This weak dependence is a 
consequence of the weak dependence of the projected clustering on apparent
\lya luminosity, as discussed in Paper II. There is also a trend that 
the two characteristic scales increase with the apparent \lya luminosity, but 
it is much weaker than that seen in Figure~\ref{fig:SB_Mh} for LBGs.

\subsection{Dependence on the Intrinsic \lya Line Profile}
\label{sec:line_prof}

\begin{figure*}
\plottwo{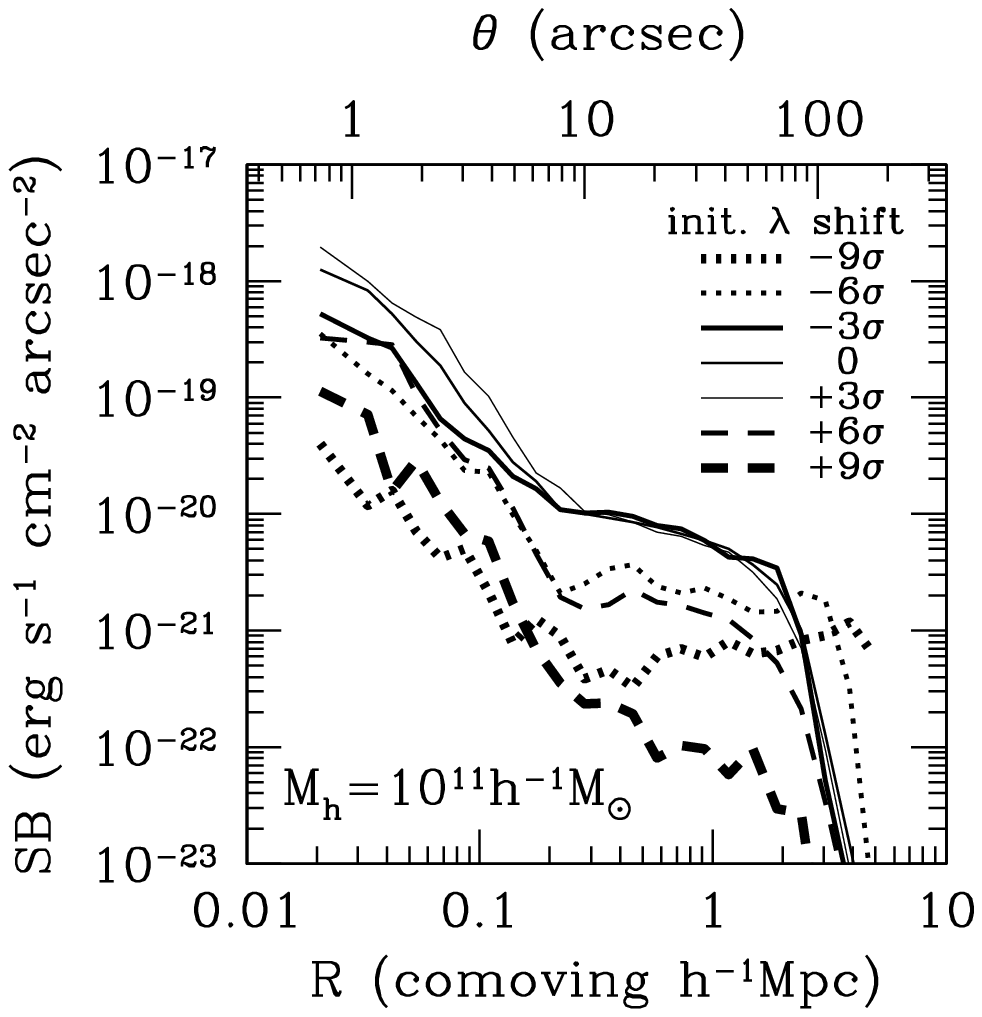}{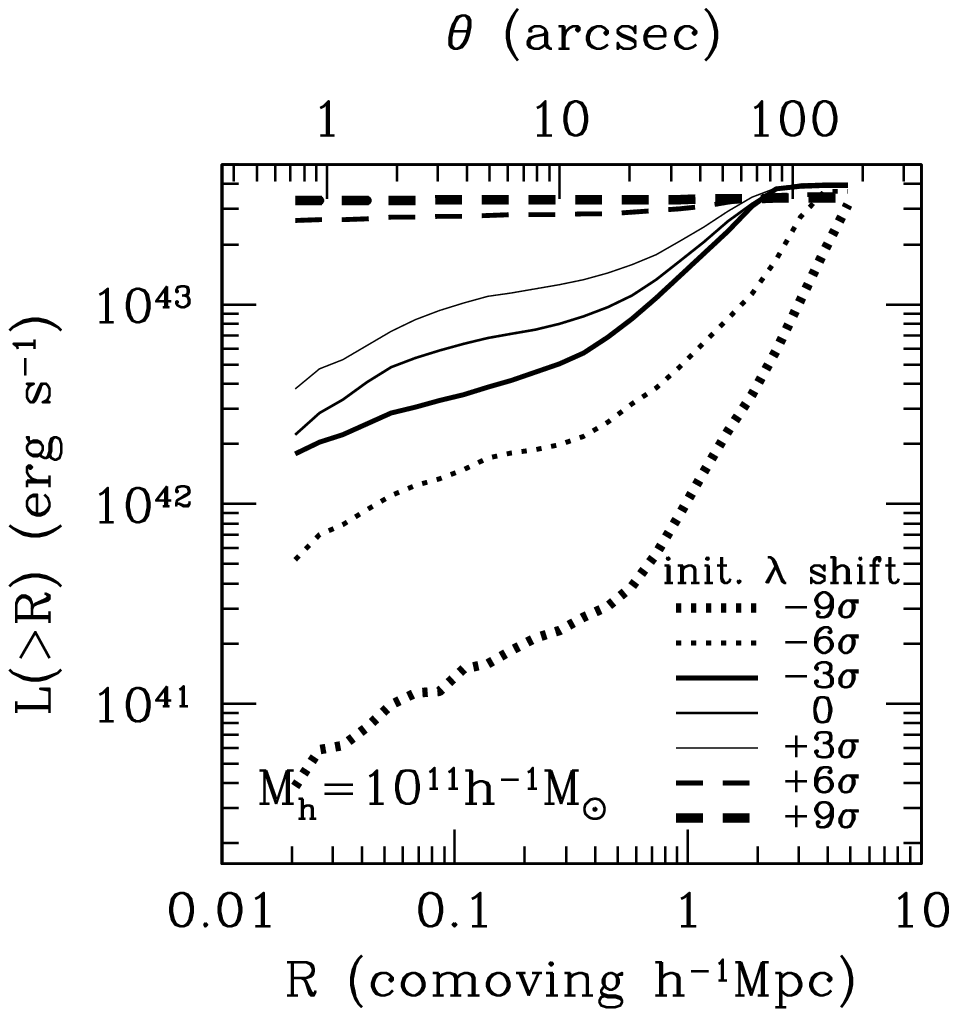}
\caption[]{
\label{fig:initfreq}
Dependence of \lya emission distribution profile on initial wavelength of \lya 
photons. The left and right panels are for the surface brightness profile and
cumulative luminosity profile, respectively. 
For each curve, the initial \lya photons are set to have a single 
wavelength, denoted in either panel as the difference from the line center in
units of the 1D thermal width of the halo gas, which is about $\sigma=69 \kms$
for $M_h=10^{11}\hMsun$. 
The initial \lya luminosity for all cases is normalized to be the same.
}
\end{figure*}

In all our discussions above, the intrinsic \lya line profile is assumed to be
Gaussian with the width determined by velocity dispersion in the halo (Paper 
I). More specifically, the standard deviation of the Gaussian profile is set 
to equal the 1D thermal velocity dispersion for gas in the halo, 
$\sigma=31.9[M_h/(10^{10} h^{-1}M_\odot)]^{1/3} \kms$. As discussed in Paper 
I and Paper II, the intrinsic line profile is one of the main uncertainties in 
our radiative transfer modeling. The spatial extent of the \lya emission 
depends on the initial line profile. 
For example, in the extreme case where all the \lya photons are
shifted far to the red from the line center within the emitting galaxy
(e.g., as a possible consequence of a strong galactic wind; 
\citealt{Dijkstra10}), they would
not be further scattered in the surrounding medium and the \lya emission
would not extend beyond the size of the galaxy.
 
To test the effect of the initial line profile on the spatial distribution of 
\lya photons, we perform radiative transfer calculations with different initial 
values of \lya wavelength for an individual source in a halo of 
$10^{11}\hMsun$ (the same source as in Figure~\ref{fig:PSF}). 
Figure~\ref{fig:initfreq} shows the \lya surface brightness and cumulative 
luminosity profiles for cases with different initial line shifts (in units of 
the thermal velocity dispersion $\sigma$ mentioned above). 

On average, bluer photons travel farther in the IGM before they redshift into 
the line center to encounter significant scatterings. Therefore, the \lya 
emission would dilute to a larger spatial extent. This is clearly seen in 
Figure~\ref{fig:initfreq} -- the spatial distribution becomes more and more 
extended as the initial \lya wavelength shift varies from $-3\sigma$ to 
$-9\sigma$. If we start with red \lya photons, the final \lya surface 
brightness profile becomes more compact. In Figure~\ref{fig:initfreq}, the 
intrinsic \lya luminosity for all curves is the same. The surface brightness 
curves for initially red photons appear to be of lower amplitude, because they 
have a central spike (within 0.02$\hMpc$) that is not shown in the left panel. 
The contribution of the central spike to the total luminosity can be figured 
out from the cumulative luminosity curves shown in the right panel. Overall, 
the \lya emission appears to be close to point-like if photons start from the 
far red side of the line center. These red photons can still have a probability
of experiencing resonant scatterings in the infall region around the halo and 
being shifted to the blue side. The surface brightness profile on scales 
$\lesssim 1\hMpc$ is expected to be contributed by these photons. The 
probability of such scatterings depends on the magnitude of the shift and the 
infall velocity around the halo. For sources in $10^{11}\hMsun$ halos, even 
with a $+6\sigma$ initial line shift (thin dashed curves in 
Figure~\ref{fig:initfreq}), the flattening part in the surface brightness 
profile can still be clearly seen (left panel), although the majority of 
the observed \lya photons are from the central region (right panel). We verify
that for a $+3.5\sigma$ initial line shift (not shown in the plot) the 
contribution from the central region to the total \lya luminosity is 
$\lesssim$40\%, less than that from the extended emission. If we take this as
a threshold for the profile to change from point-like to extended, it would
correspond to $\sim 250\kms$.

If the wavelength shift of the initial \lya photons is in the range of 
$\pm 3\sigma$, the resultant surface brightness distribution is not 
sensitive to the initial \lya wavelength, as shown by the solid curves in
Figure~\ref{fig:initfreq}. 
This is easy to understand: photons starting at a wavelength
shift $\pm 3\sigma$ undergo many scatterings in the core, leading to a
distribution that loses memory of the original wavelength and to a
nearly constant surface brightness profile. But photons with a larger
initial wavelength deviation are not scattered enough times to be
redistributed in this way before they escape. Therefore, the predicted
\lya profile should be reliable except when many photons are emitted
from the central galaxy with a wavelength shift of several times the
velocity dispersion.

To eliminate most of the extended \lya emission around the galaxies, 
the initial line shifts (after photons escape the ISM) need to be 
larger than $250\kms$ for halos of $10^{11}\hMsun$. Galactic winds can have 
the effect of shifting \lya emission toward red 
\citep[e.g.,][]{Verhamme06,Dijkstra10}. 
From observations of $z\sim 3$ galaxies, \citet{Steidel10} find that the
interstellar absorption line shifts with respect to the H$\alpha$-defined 
galaxy systemic redshift by $165 \pm 140 \kms$ (mean $\pm$ standard 
deviation) for galaxies of baryon mass (stars plus cold gas) ranging from
$\sim 1\times 10^{10}\Msun$ to $\sim 3\times 10^{11}\Msun$. The uncertainty 
of velocity measured for individual galaxies is estimated to be 
$\sim 130\kms$. If the wind is confined in a region not far from the ISM, the
above observational results indicate that the initial shift in \lya emission 
could be above $\sim 300\kms$. However, even in such a case, the shift may not 
be sufficient for \lya photons to completely decouple from the circumgalactic 
and intergalactic environments, depending on the spatial distribution of the 
wind. 
Observationally, the outflowing neutral gas of the wind is not isotropic, 
usually collimated in a bipolar fashion
\citep[e.g.,][]{Bland88,Shopbell98,Veilleux02}. It is likely that \lya 
emission escaping in directions other than the wind would gain only a smaller 
redward shift, and the source would still become extended in the end.
The wind effect on the spatial distribution of \lya emission is also expected 
to be less significant for sources before reionization than those after 
reionization, since the higher neutral density boosts the scattering optical 
depth.

It is also possible that the wind could reach scales much larger than the ISM
\citep[e.g.,][]{Steidel10}. In such a case, the initial line shifts
(after photons escape the ISM) is not much affected by the wind, but the wind
alters the distribution and dynamics of neutral gas in the circumgalactic
region and therefore the details in the \lya radiative transfer process. While
the surface brightness profile of \lya emission may change accordingly, the
extended \lya emission should remain as a generic feature. Such a possibility
deserves further investigations using high resolution galaxy formation 
simulations with a wind prescription.

A change in the initial \lya line profile can lead to a change in the 
observed \lya luminosity and therefore the observed \lya luminosity 
function. However, as discussed in Paper I,  a combination
of changes in the star formation prescription used, the  stellar IMF adopted, 
and the dust extinction is able to keep the predicted \lya luminosity function
to match the observed one. On the other hand, the change in the initial \lya 
line profile can alter the strength of the coupling between the observed \lya 
emission and the circumgalactic and intergalactic environments. This 
would show up in the clustering of LAEs (Paper II), which provides an
interesting way to constrain the initial \lya line profile and the effect of 
galactic wind.

\subsection{Effect of Grid Resolution}
\label{sec:resolution}

\begin{figure}
\plotone{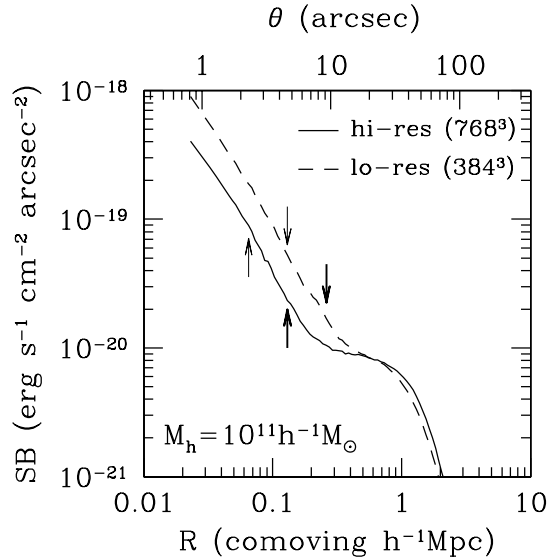}
\caption[]{ 
\label{fig:resolution} 
Effect of the resolution of the grid on the \lya surface brightness profile.
Shown here are average (one-halo term) profiles from stacking narrowband images 
of individual sources residing in halos of $10^{11}\hMsun$. The solid curve
is obtained by performing \lya radiative transfer calculation with gas 
properties (density, temperature, and peculiar velocity) represented with 
a 768$^3$ grid (default case in this paper), while the dashed curve shows
the case with a lower resolution grid (384$^3$). For each case, the thin 
and thick arrows mark the half and full size of a grid cell, respectively. 
Note that the image pixels to collect \lya photons is much finer than the grid 
cell. Calculation with the higher resolution grid tends to predict more 
extended \lya emission.
}
\end{figure}

For the main results in this paper, our \lya radiative transfer calculation
is performed in a $100^3 \VhMpc$ simulation box, where gas properties 
(density, temperature, and peculiar velocity) are sampled with a 768$^3$ 
grid. All \lya photons are originally emitted from point sources and the
scatterings of \lya photons inside individual cells are also followed. 
As shown in Paper I, the dynamics of the infall region around halos plays
a significant role in determining the distribution of \lya photons, so we
expect that the grid resolution in our calculation is able to predict
the generic features in the extended \lya emission. We perform a resolution
test to show that this is the case.

It would be ideal to do such a test with a higher resolution grid. 
However, the available reionization simulation outputs prevent us from doing 
this. Therefore, we degrade the default 768$^3$ grid used in this paper to a 
lower resolution 384$^3$ grid by binning the gas density, temperature, and 
peculiar velocity. We then perform the \lya radiative transfer calculation
with such a grid for $10^{11}\hMsun$ halos and compare the resultant average
\lya surface brightness profile (the one-halo term) with that with the 
$768^3$ grid.

The comparison is shown in Figure~\ref{fig:resolution}. For each resolution 
case, the thin and thick arrows mark the half and full size of a grid cell, 
respectively. For a source located at the center of a cell, the initial
\lya photons see a uniform gas distribution within a radius of about half 
cell size. However, the final \lya surface brightness distribution inside
this resolution radius is determined by the gas properties not only in
this central cell but also in the surrounding cells. The reason is that
\lya photons propagating into surrounding cells have chances to be scattered 
back into the central cell. That is, the surface brightness profile is not
a local quantity as a result of radiative transfer. Therefore, the small scale
surface brightness profile (in particular the slope) is not completely a 
result of the grid resolution, as indicated by the fact that the powerlaw-like
inner profile naturally extends to scales a few times larger than the 
resolution scale. The transition to a plateau is not an artifact of grid
resolution, either.

With the low resolution grid, the general features in the \lya surface 
brightness profile remain the 
same, a steep drop on small scales and an extended plateau on large 
scales before a sharp cutoff. However, the amplitude of the profile on small
scales is about a factor of two higher with the low resolution grid, and it is
compensated by a slightly lower amplitude on large scales. Therefore, a lower
resolution has the effect of making the emission more concentrated.

If the grid were a factor of two finer than the default grid, we expect the
predicted small scale surface brightness profile to be lower and the 
difference should be smaller than a factor of two. The \lya emission would 
become {\it more} extended. We therefore expect that the generic prediction 
of the extended \lya emission would be strengthened with an improved grid 
resolution. 

Even though for the one-halo term of the surface brightness profile the
overall cusp-to-plateau behavior does not depend on the grid resolution, 
the transition scale decreases for a higher resolution. 
Figure~\ref{fig:resolution} seems to show that the transition happens at
a scale about twice the cell size, indicating a resolution effect. Although 
cells around the source cell contribute to the determination of the profile of
the central cusp, the cuspy profile itself is inevitably expected to also 
depend on the exact density, temperature, and velocity distributions of 
neutral gas within the source cell. The effects discussed in 
\S~\ref{sec:line_prof} add further uncertainties in the cuspy profile.

Despite the above uncertainties, our prediction on the existence of the 
extended \lya emission around high redshift star-forming galaxies remains 
valid, as long as \lya photons encounter scatterings in the circum-galactic and
inter-galactic media. The predicted two characteristic scales in the surface
brightness profile from stacked images should also remain valid. The outer 
scale, which is mainly determined by the one-halo plateau region, is less
sensitive to the exact central cuspy profile. The inner transition scale, on
the other hand, depends on the amplitude and slope of the central cusp, 
which may not be properly resolved in our model.
If the grid resolution is the only source of uncertainty, the inner 
transition scale of 0.1--0.2 Mpc (comoving) seen in the stacked profiles in 
\S~\ref{sec:extended} should be regarded as upper limits. 

\section{Discussion}

The \lya emission from star-forming galaxies is extended because
of resonant scattering in the circumgalactic and intergalactic environments. In
the stacked narrowband \lya image, the change from a declining to nearly flat 
surface brightness profile at a characteristic scale of a few tenths of Mpc
(possibly an upper limit, see \S~\ref{sec:resolution}).
and that from the flat profile to a declining profile at a few Mpc are
evidence for the resonant scattering nature of the extended \lya emission.

According to our radiative transfer modeling, the visibility of \lya emission
from star-forming galaxies depends on the circumgalactic and intergalactic
environments around sources, and only the central, high surface brightness
part can be detected as LAEs. The extended \lya emission around the 
star-forming galaxies is predicted to exist (as long as \lya photons escape 
from the ISM), no matter whether they can be detected as LAEs or not. This 
implies that the extended emission can be detected from both LBGs and LAEs. 
In practice, to detect the extended \lya emission around LBGs, we need to know 
their redshifts and take their \lya narrowband images. Extended \lya
emission from LAEs is more readily detected, since LAEs are typically
discovered in narrowband surveys.

Our radiative transfer modeling is performed for $z=5.7$ star-forming galaxies.
The generic picture should remain valid at other redshifts -- as long as \lya 
photons encounter significant scatterings in the circum-galactic/intergalactic 
media, the \lya emission would become extended. However, the physical 
conditions in and around star-forming galaxies evolve with redshift. For 
example, compared to $z=5.7$ galaxies, we expect a lower neutral hydrogen 
density around $z\sim 3$ galaxies, as a result of the expansion of the
universe and the evolution of the UV background. The Hubble expansion
rate decreases, a change that may partially compensate the decrease in 
neutral density for \lya scattering. The extended \lya emission (e.g., the
surface brightness profile) at $z\sim 3$ may not follow the details of our 
$z=5.7$ 
prediction, although it is expected to show up. We reserve the investigation 
of the redshift dependence of the extended \lya emission for future work. 
We do not include dust in our model. The effect of ISM dust on the \lya line 
depends on whether dust distribution is clumpy or not
\citep{Neufeld91,Hansen06}. Optical, UV, and \lya observations of local 
star-forming galaxies show evidence that ISM kinematics and geometry play 
a more significant role than dust in affecting the \lya emission 
\citep[e.g.,][]{Giavalisco96,Keel05,Atek08,Atek09}. We also expect the dust
content to decrease towards high redshift (e.g., $z\sim 6$), and there is
observational evidence for low dust content in $z\gtrsim 6$ galaxies (e.g.,
\citealt{Bouwens10,Ono10}). It is likely,
however, that the existence of dust can lead to variations in the amount 
of \lya photons escaping from the ISM, and thus add additional scatter in the 
extended \lya emission among individual galaxies.
Keeping all the above caveats in mind, we discuss the current status of 
observing the extended \lya emission.

Observations have long been suggesting the existence of extended \lya 
emission. Based on {\it Hubble Space Telescope (HST)} broad-band and 
{\it New Technology Telescope (NNT)} narrow-band images,  \citet{Moller98} 
found that three LBG-like galaxies near a $z=2.8$ quasar are more extended in
\lya emission (half-light radius of 0.5\arcsec--0.7\arcsec) than in the 
continuum (half-light radius of 0.13\arcsec). Observations with NNT and  
Very Large Telescope (VLT) of a larger sample of LAEs ($\sim 70$) 
in quasar fields at $z\sim 3$ further strengthen the case for extended \lya 
emission
(e.g., \citealt{Fynbo01,Fynbo03}; also see \citealt{Laursen07} for comparing
one of their objects with the results from a model with \lya radiative 
transfer). In particular, \citet{Fynbo01} found the
size of \lya emission to be 0.65\arcsec ({\it FWHM}) in the stacked 
narrow-band images of seven LAEs. Although the spatial extent of the 
\lya emission from the above observations is small compared to what we 
predict, owing to the surface brightness detection limit, it is clear that 
\lya emission is more extended than the underlying UV emission.

There are efforts to use stacked narrowband images in deep surveys to study 
the compactness of the \lya emission. \citet{Hayashino04} find that 24 
spectroscopically
confirmed $z\sim 3$ LBGs in \citet{Steidel03} fall within the redshift range 
of their deep narrowband image observed with Subaru telescope (with only two 
of them being identified as LAEs). 
The UV luminosity of these LBGs ranges from 
-20 mag to -22.5 mag (based on the observed ${\cal R}_{\rm AB}$ magnitude;
\citealt{Steidel03}).
After removing two gigantic \lya blobs, the composite narrowband 
(continuum subtracted) image of the 22 LBGs shows \lya emission that extends
at least to $5\arcsec$ (about $0.1\hMpc$ comoving; see their Fig.8). The 
surface brightness drops from $\sim 3\times 10^{-18}\SBunit$ at 
$\sim 1\arcsec$ to 
a few times $\sim 3\times 10^{-19}\SBunit$ at $\sim 4\arcsec$, corresponding
to a power-law index of $\sim -1.7$. They further note that 13 LBGs with no
individual detection of extended \lya emission also show extended \lya 
emission on their composite image. The findings of \citet{Hayashino04} are
in line with our prediction that extended \lya emission exists for LBGs, and
the slope of the surface brightness profile is also in broad agreement with
our prediction for UV bright sources. 

\citet{Steidel11} present the result of stacking analysis of deep narrow-band
observation of 92 UV continuum-selected star-forming galaxies at 
$2.2<z<3.1$.
\footnote{ 
It is exciting that our generic prediction (if dated from the initial 
submission of our paper) started to be tested or verified in merely three 
months, although many details remain to be investigated.
}
Some sources are the same ones as in \citet{Hayashino04} but with much deeper
narrow-band observation. The stacked narrow-band image shows diffuse \lya 
emission around star-forming galaxies, with surface brightness drops from
a few $\times 10^{-18}\SBunit$ at the center to $\sim 10^{-19}\SBunit$ at a 
radius of $\sim 10\arcsec$ ($\sim$80 proper kpc). The extended emission exists
in the stacked image of sub-samples -- LBGs, LAEs, and \lya absorbers all
show qualitatively similar diffuse emission. The overall surface brightness 
amplitude of the extended emission increases with the central \lya emission.
Although the redshift difference prevents us from a direct comparison between 
the observation and our prediction, the detected diffuse \lya emission supports
our generic picture that resonant scatterings in the circum-galactic and 
intergalactic media causes \lya emission originated from star formation in 
the central galaxy to become extended. If we make a naive comparison, the 
\lya emission is detected up to the inner characteristic scale in the surface 
brightness profile (e.g., Figures~\ref{fig:SB_Mh} and \ref{fig:SB_Lap}), not 
reaching the plateau yet. In \citet{Steidel11}, the clustered sources are 
intentionally masked before image stacking, which may change the shape of 
plateau if it could be detected. We plan to extend our model to $z\sim 3$ 
star-forming galaxies and make direct comparison with the observation in 
\citet{Steidel11}.

\citet{Ono10} use stacked multiband images of LAEs at $z\sim$5.7 and 6.6 in 
the SXDS to investigate the stellar population 
of LAEs. For the narrowband stacked images, they find that the \lya fluxes 
level off for apertures larger than $5\arcsec$ in diameter. Such a behavior
in the surface brightness profile is predicted in our model as the transition
to the two-halo term dominated regime. It is likely that the feature seen by
\citet{Ono10} is caused by the clustering of \lya emitting sources.

\citet{Finkelstein11} detect \lya emission from three spectroscopically 
confirmed $z=4.4$ LAEs in the HST/ACS F658N 
narrowband imaging data. They find that \lya emission appears to be more 
extended than the UV emission. However, the half-light radii of \lya emission 
are quite small, in the range of 0.1$\arcsec$--0.2$\arcsec$. One thing to 
notice is that the observation is not deep enough to reach the sensitivity of 
detecting the much larger extended \lya halo predicted by our model. This is 
clear from the fact that the surface brightness level in the {\it HST} 
narrowband observation turns out to be above a few times $10^{-16}\SBunit$. It 
is also supported by 
the fact that the three objects have significantly higher \lya fluxes detected 
from ground-based narrowband image (with larger telescope apertures and reduced
sensitivity to read noise; \citealt{Finkelstein11}). With {\it HST}/WFPC2 
narrowband images of eight $z=3.1$ LAEs, \citet{Bond10} find half-light radii
of \lya emission to be in a similar range, 0.15$\arcsec$--0.30$\arcsec$. The 
surface brightness level (a few times $10^{-17}\SBunit$) reached by the 
observation is still too high to probe the extended \lya emission predicted 
by our model.

Extended \lya emission can also be detected through deep spectroscopic 
observations. \citet{Rauch08} conduct a long-slit search for $2.67<z<3.75$ 
low surface brightness \lya emission from a 92-hour long exposure with the
VLT, reaching a $1\sigma$ surface brightness detection limit of 
$8\times 10^{-20}\SBunit$ in a 1 arcsec$^2$ aperture. They find 27 objects
with possible \lya emission lines. These objects are typically fainter than 
the LAEs and LBGs in \citet{Steidel11}. About one third of them show clear 
signatures of extended emission.  If the surface brightness profile at large 
radii is approximated by a power-law, the slope is generally in the range of 
$-1$ to $-3$ (their Figure 17). 
In the stacked surface brightness profile (their Figure 20), the emission 
clearly extends to $\gtrsim4\arcsec$ (about comoving 90$\hkpc$), dropping from 
$\sim 8\times 10^{-19} \SBunit$ at a radius of 1\arcsec  to 
$\sim 1\times 10^{-19} \SBunit$ at large radii. The profile roughly follows
a power law with an index in the range of $-1.5$ to $-2.0$. This is in the 
ballpark of the predicted extended emission, although a direct comparison
is not possible because of the redshift difference. The spatially resolved 
spectra of the extended emission remain as an interesting aspect to be 
explored in our model.

The extended \lya halo from \lya scatterings predicted by our model shares
some similarities with the one around a quasar before reionization described
in \citet{Loeb99}. Unlike \citet{Loeb99}, who assume a uniform, zero 
temperature IGM undergoing Hubble expansion, we use a realistic distribution
of gas density, temperature, and velocity around a star-forming halo from
cosmological reionization simulation. We expect that extended
\lya emission around star-forming galaxies also exists before reionization, 
which merits further study. However, the \lya surface brightness is
expected to be proportional to the source luminosity and to not depend on
the neutral fraction in the IGM as long as the majority of photons are
scattered, while it should decline with redshift as $(1+z)^{-4}$, making
the detection increasingly difficult at higher redshift.

Our radiative transfer calculation is done for sources residing in halos
above $5\times 10^9\hMsun$. Star formation in lower mass halos, down to
$\sim$$2\times 10^8\hMsun$ at $z\sim 5.7$ (with virial temperature 
$>$$10^4K$ for gas to cool; see \citealt{Trac07}), can also contribute to the 
two-halo term of the stacked profile. According to \citet{Trac07} (see their 
Fig.1), the total star formation rate in $<$$5\times 10^9\hMsun$ halos can  
contribute about two thirds of the global rate. 
The amplitude 
of the clustering with these low mass halos is lower, and we expect that
including the \lya emission from the $<$$5\times 10^9\hMsun$ halos would 
boost the two-halo term of the surface brightness profile presented in this 
paper at the factor of two level and that the two characteristic scales would 
shift inward.
Our model does not resolve subhalos,
but we do not expect a large subhalo (or satellite galaxy) fraction for 
galaxies at redshift $z\sim 6$. 
If we consider a mass threshold sample of halos
with $M_h>M_{\rm min}$ and assume that the mean number of subhalos increase 
linearly with halo mass as $M_h/(10M_{\rm min})$ \citep{Kravtsov04}, we find 
that the satellite fraction is about 15\% for $M_{\rm min}\sim 10^{11}\hMsun$.
As subhalos are distributed inside virial radius of the parent halo, including 
star formation in subhalos would make the central cuspy surface brightness 
profile in the stacked image more extended, which would smooth the transition 
to the plateau part of the profile and increase the inner characteristic scale.
Investigating the exact magnitude of the effect is beyond the scope of this 
paper.

Besides star formation, there are other sources of \lya emission that 
increase in the dense regions around galaxies, including fluorescence from 
the UV background and cooling radiation. At $z\sim 5.7$, the fluorescent 
emission caused by the UV background is expected to be at the level of 
a few times $10^{-21}\SBunit$ \citep[e.g.,][]{Gould96,Kollmeier10}, much 
fainter than the signal of interest in this paper. Cooling radiation may have
a noticeable contribution to the stacked \lya surface brightness profile, 
but there are uncertainties in the theoretical prediction 
\citep[e.g.,][]{Faucher10}. Further work is needed to see whether \lya emission
from cooling radiation can change the profile and to see how one can separate 
\lya emission caused by star formation from other sources of \lya emission.

Finally, for observational efforts, sky subtraction is an important step in 
revealing the extended \lya emission from star-forming galaxies. 
To compare the observation with the theoretical prediction from simulations, 
a detailed description of the method of sky subtraction used in the 
observation is necessary, which should be reproduced in the same way in 
the model.

\section{Conclusion}

After escaping from the ISM, \lya photons converted from ionizing photons in 
star-forming galaxies experience scatterings in the circumgalactic and 
intergalactic media. Such a radiative transfer process makes \lya emission
from these galaxies spatially extended. In this paper, built on the radiative
transfer modeling of LAEs in \citet{Zheng10} and \citet{Zheng11}, we 
investigate the predicted spatial distribution of \lya emission that can
be measured from the stacked narrowband image of these galaxies.

In general, the predicted surface brightness profile measured from the 
stacked image has two characteristic scales, an inner one at tenths of Mpc and
an outer one at about 1 Mpc. The profile shows a central cusp inside the inner 
scale, an approximate plateau between the two scales, and an extended tail 
beyond the outer scale. 
The inner scale may possibly be an upper limit, as an effect of the grid 
resolution in the simulation (see \S~\ref{sec:resolution}).

The stacked surface brightness profile can be understood as a superposition
of the brightness distribution from the stacked sources themselves (one-halo
term) and that from neighboring clustered sources (two-halo term).
The two-halo term is the profile of the (angular) two-point correlation
function (plus one) smoothed by the extended \lya emission profile (PSF) of
individual sources. The smoothing makes the profile flattened on scales
smaller than the spatial extent of \lya emission of clustered sources.
The outer characteristic scale in the stacked surface brightness profile marks 
this spatial extent, and the plateau in the profile is a consequence of the
smoothing effect. The transition from one-halo term domination to two-halo
term domination leads to the inner characteristic scale seen in the stacked 
surface brightness profile.

For continuum selected galaxies (LBGs), the amplitude of the stacked \lya 
surface brightness profile increases with the source UV luminosity (or halo
mass) on both small and large scales. The two characteristic scales also 
increase with the UV luminosity. The central cusp becomes steeper for higher 
UV luminosity.  For \lya line selected galaxies (LAEs), the amplitude of the 
central cuspy profile increases with observed \lya luminosity and the slope is 
steeper than that of LBGs. Beyond the inner characteristic scale, the 
amplitude of the surface brightness profile only has a weak dependence on 
observed \lya luminosity.  The inner and outer characteristic scales do not 
show strong dependence on \lya luminosity, either.

Because of the contribution from source clustering to the stacked surface 
brightness profile, the cumulative \lya luminosity from the stacked image
does not converge to the intrinsic \lya luminosity of the stacked sources. 
It is therefore not straightforward to estimate the total \lya emission from 
the stacked image. 

In detail, the surface brightness profile depends on the initial line profile 
of \lya emission (after \lya photons escape the ISM). The source would appear
to be more compact in the narrowband image as the initial line shift 
toward red increases. If this shift is caused by galactic
wind, the wind velocity needs to be comparable to a few times the virial 
velocity of the host halo and the wind needs to be largely isotropic to 
make \lya emission point-like. Conversely, from the
measured \lya surface brightness profile in the narrowband image, it may be
possible to constrain the effect of galactic winds. It is worth further study
along this line.

Our particular prediction here is for 
sources after reionization is complete ($z\sim 5.7$). 
It is interesting to investigate how the prediction changes for 
star-forming galaxies at lower redshifts, because of the large amount of 
efforts in observing star-forming galaxies around redshift 2--4 (e.g., 
\citealt{Rauch08,Steidel11}) and in the local universe (e.g., 
\citealt{Kunth03,Hayes07,Ostlin09}). It is also necessary to study
the extended emission for sources at higher redshifts, before reionization
is complete, given the increasing observational efforts (e.g., 
\citealt{Ouchi10,Kashikawa11}).

At the time of our initial submission, we conclude that
deep narrowband photometry from large ground-based telescopes is on the verge 
of detecting the extended \lya emission around star-forming galaxies, including
LBGs and LAEs \citep[e.g.,][]{Hayashino04,Ono10}. 
Now the latest observation by \citet{Steidel11} indeed revealed the extended 
\lya emission to large radii ($\sim 10\arcsec$) around LBGs and LAEs, 
starting to verify our prediction and 
provide stringent test to the theoretical model. 
The detection of the predicted \lya emission supports the generic picture that
\lya radiative transfer in the circumgalactic and intergalactic environments
produces extended \lya emission. 
The extended \lya emission opens a new 
window to study the circumgalactic and intergalactic environment of 
high-redshift star-forming galaxies. The surface brightness profile encodes
information about cold baryons around galaxies, including their density,
temperature, and velocity. All of these properties could be modified by 
galactic wind, so the extended emission in principle can put constraints on
the galactic wind. The stacked \lya image also includes 
contributions from faint galaxies that cannot be detected in a single
exposure, which provides an opportunity to study low luminosity 
star-forming galaxies. The extended emission gives us a better idea on the
total amount of \lya emission from galaxies. When compared with the UV 
emission or other optically-thin line emission (e.g., H$\alpha$ emission), 
we may infer the dust distribution and its effect on \lya photons.
With integral-field-units observations of high-redshift star-forming 
galaxies, we would have stacked spectra for the extended emission and
expect to learn more about galaxy environments. Details on how to extract
all the information encoded in the extended \lya emission need to be 
investigated, and \lya radiative transfer 
calculation, in combination with sophisticated models of star-forming galaxies 
and their environments, will play an irreplaceable role.

\acknowledgments

ZZ thanks Masami Ouchi, Tomoki Saito, and Chuck Steidel for helpful 
discussions and the 
hospitality of the Institute for the Physics and Mathematics of the Universe.
We thank Andy Gould for useful comments and the referee for constructive 
suggestions.
ZZ gratefully acknowledges support from Yale Center for Astronomy and
Astrophysics through a YCAA fellowship. ZZ and DHW thank the Institute for 
Advanced Study for their hospitality during part of the work. RC acknowledges
support from NASA grants NNG06GI09G and NNX08AH31G. JM is supported by the 
Spanish grant AYA2009-09745.

{}

\end{document}